\documentclass[aps,preprint,prd]{revtex4}
\usepackage{amssymb}
\usepackage{amsmath}
\usepackage{graphicx}
\usepackage{fancyhdr}
\newcommand{\bea}{\begin{eqnarray}}
\newcommand{\eea}{\end{eqnarray}}

\begin{document}

\title{How the Inverse See-Saw Mechanism Can Reveal Itself Natural, Canonical and Independent of the Right-Handed Neutrino Mass}
\vspace{2.5cm}

\author{A. G. Dias$^a$\footnote{email: \tt
alex.dias@ufabc.edu.br}, C. A. de S. Pires$^b$\footnote{email: \tt
cpires@fisica.ufpb.br}, P. S. Rodrigues da Silva$^b$\footnote{email: \tt
psilva@fisica.ufpb.br}}

\affiliation{{ $^a$Centro de Ci\^encias Naturais e Humanas, UFABC, Santo Andr\'e, SP,  Brasil\vspace{0.25cm}\\}}

\affiliation{{ $^b$Departamento de
F\'{\i}sica, Universidade Federal da Para\'\i ba, Caixa Postal 5008, 58051-970,
Jo\~ao Pessoa, PB, Brasil}}

\date{\today}

\begin{abstract}
The common lore in the literature of neutrino mass generation is that the canonical see-saw mechanism beautifully offers an explanation for the tiny neutrino mass but at the cost of introducing right-handed neutrinos at a scale that is out of range for the current experiments. The inverse see-saw mechanism is an interesting alternative to the canonical one once it leads to tiny neutrino masses with the advantage of being testable at TeV scale. However, this last mechanism suffers from an issue of naturalness concerning the scale responsible for such small masses, namely, the parameter $\mu$ that is related to lepton number violation and is supposed to be at the keV scale, much lower than the electroweak one. However, no theoretical framework was built that offers an explanation for obtaining this specific scale. In this work we propose a variation of the inverse see-saw mechanism by assuming a minimal scalar and fermionic set of singlet fields, along with a $Z_5\otimes Z_2$ symmetry, that allows a dynamical explanation for the smallness of $\mu$, recovering the neat canonical see-saw formula and with right-handed (RH) neutrinos free to be at the electroweak scale, thus testable at LHC and current neutrino experiments.
\end{abstract}

\maketitle

\section{ Introduction}
\label{intro}
If, on one hand, the experiments on neutrino oscillation confirm that neutrinos are massive particles~\cite{experiments}, on the other hand,  no definite understanding of the inferred neutrino mass scales and mixing emerging from these experiments is available yet. The see-saw mechanism is considered the most elegant scheme for explaining the smallness of the neutrino masses.  The essence of the  mechanism lies in the fact that lepton number must be violated in some way at a high energy scale.
Basically, the outcome of this mechanism is the canonical see-saw formula for the neutrino masses, $m_\nu = \frac{v^2_w}{M}$, where $v_w$ is the vacuum expectation value (VEV) of the Higgs field in the electroweak scale and $M$ is the high energy scale associated to new physics where lepton number violation occurs. For $v_w\approx 10^2$~GeV, a sub-eV mass scale naturally arises if $M$ can be associated with some grand unifying theory.
There are at least three different and independent ways of realizing this. The simplest realization is via the existence of heavy RH neutrinos (with a Majorana mass term) coupled to the left-handed (LH) ones through the usual scalar doublet of the electroweak standard model (SM). This see-saw mechanism is called in the literature  "type I see-saw mechanism"~\cite{seesawI}. It can also be realized via the existence of an extra triplet of scalars or fermions, in which case the mechanism is called  "type II"~\cite{seesawII} and "type III  see-saw mechanism"~\cite{seesawIII}, respectively.  In spite of providing a simple explanation for the smallness of the neutrino masses, all these three  mechanisms lack  being phenomenologically testable once the new physics phenomena engendered by them take place at a too high energy scale, not accessible to experiments.

As the large hadron collider (LHC) is probing the physics at TeV scale, a plethora of new see-saw mechanisms working at TeV scale were proposed in the last ten years~\cite{alternativeseesaw}.  However, the first see-saw mechanism, whose new physics should manifest at this scale dates back to the 1980s, which is nowadays called  the "inverse see-saw" (ISS) mechanism~\cite{inverseseesaw}. It is a different realization of the see-saw mechanism in the sense that  the smallness of the neutrino masses is due to the violation of lepton number at a low energy scale and, as a by-product, it predicts RH neutrinos at TeV scale. Recently, with the advent of the LHC,  it has received enough attention in various contexts as gauge extensions of the standard model, supersymmetry, grand unified theories, etc~\cite{recentdevelopment}.

General analysis concerning the mass matrix, mixing, and unitarity violation applicable for the ISS mechanism  was done in Refs.~\cite{gl2000,hlr2011}. We shall consider here the ISS mechanism realization, assuming that there are six RH neutrinos which are singlets under the SM symmetry group. Three of them, which we denote $\nu_{\alpha R}$, $\alpha =1, 2, 3$, are the Dirac partners of the known LH neutrinos, $\nu_{\alpha L}$. The remaining   three are new RH neutrinos which are simply referred  as $N_{i R}$, $i =1, 2, 3$. The  mass matrix, which is constructed from bilinear terms \, left after symmetry breaking, involving all these neutral fermions is supposed to have the  following texture: 
\begin{equation}
M_\nu=
\begin{pmatrix}
0 & m^T_D & 0 \\
m_D & 0 & M_N^T\\
0 & M_N & \mu
\end{pmatrix},
\label{ISSmatrix}
\end{equation}
in the basis  $(\nu_L\,,\, \nu^c_L\,,\,N^c_L)$, where we are using the notation of charge conjugation $\nu^c_L\equiv (\nu_R)^c$ and $N^c_L\equiv (N_R)^c$ .

The mass Lagrangian leading to the mass matrix in  Eq.~(\ref{ISSmatrix}) is
\begin{equation}
{\cal L}=-\overline{\nu_R} m_D \nu_L  - \overline{N_R} M_N \nu^c_L - \frac{1}{2} \overline{N_R} \mu N^c_L + {\mbox h.c.}\,.
\label{massterms}
\end{equation}
$m_D$, $M_N$  and $\mu$ are $3\times 3$ mass matrices. Without loss of generality,  we consider that $\mu$ is diagonal. It is also supposed that there is a hierarchy for the typical mass scales in the matrices so that $\mu << m_D << M_N$. What makes the texture  in Eq.~(\ref{ISSmatrix}) interesting from the phenomenological point of view can be seen directly after block diagonalization of $M_\nu$ which provides, in a first approximation, the following effective neutrino mass matrix for the standard neutrinos:
\begin{equation}
m_\nu = m_D^T M_N^{-1}\mu (M_N^T)^{-1} m_D.
\label{inverseseesaw}
\end{equation}
The double suppression by the mass scale connected with $M_N$ makes it possible to have such a scale much below that one involved in the canonical see-saw. It happens that standard neutrinos with mass at the sub-eV scale are obtained for $m_D$ at the electroweak scale, $M_N$ at the TeV scale, and $\mu$ at the keV scale. In this case, all the six RH neutrinos get masses at the TeV scale and their mixing with the standard neutrinos is modulated by the ratio $M_DM_N^{-1}$. The core of the ISS is that the smallness of the neutrino masses is guaranteed by assuming that the $\mu$ scale is small, and in order to bring the RH neutrino mass down to the TeV scale, it has to be at the keV scale.

The appealing feature behind this mechanism is that it implies RH neutrino masses at the TeV scale, which can possibly be probed at LHC~\cite{probedLHC} and future long-baseline neutrino experiments, making it a testable see-saw mechanism. That is a huge gain, but one could certainly bring back the question of naturalness now disguised in the smallness of the $\mu$ parameter, leaving the impression that we just traded the unnatural Yukawa couplings by an unnatural mass scale. In the ISS, it is argued that such a scale is naturally small in the sense it comes from a slight violation of the lepton number symmetry, $U(1)_L$, but with no dynamical reason for this. In our view, it seems imperative to find out a way to justify the smallness of the $\mu$ parameter if we want to keep the elegance and accessibility of the ISS mechanism. That is the contribution we want to add to this intriguing question in this work.

To this end, we propose a modification of the original ISS mechanism, with the aim of  generating the canonical see-saw formula for the neutrino masses, by providing a field theoretical framework to obtain a naturally small $\mu$ scale~\cite{otherproposal}. In what follows, we present, in general terms, the idea of the mechanism, and develop a simple extension of the SM which realizes such a proposal.

\section{The new mechanism}
\label{sec:newmechanism}
In order to present the idea of a natural ISS mechanism, we assume that, in Eq.~(\ref{ISSmatrix}), $m_D $ is connected to  $v_{w}$ and that  $M_N$ is determined by another VEV, $v$. In this way, the order of magnitude involved  in Eq.~(\ref{inverseseesaw}) is such  that $m_\nu\propto\frac{v^2_w}{v^2}\mu$.  Let us consider  that   $\mu$ originates from an Yukawa term, $\lambda \sigma^0 \overline{N_R} (N_R)^C$, where $\sigma^0$ is a heavy scalar singlet  that carries two unities of lepton number, so as to conserve lepton number at this level. Thus, when $\sigma^0$ develops a VEV (we call it $v^{\prime}$), we can identify $\mu=\lambda v^{\prime}$. Now comes the crucial assumption of this work.  Unlike the original ISS mechanism where lepton number is explicitly violated by a Majorana mass term for the RH neutrinos, $N_R$, in our mechanism, it is assumed that explicit lepton number violation occurs only through the scalar potential of the model. The point is that if we choose an appropriate set of fields and symmetries, our scalar potential might possess a minimum that constrains the parameters of the model (the tadpole equation for the singlet scalar $\sigma^0$) as to provide $v^{\prime}=\frac{v^2}{M}$, where $M$ represents the scale where lepton number is explicitly violated. In this case, we obtain $\mu=\lambda \frac{v^2}{M}$. Notice that if $M\approx 10^{13}$~GeV  and $v\approx 1$~TeV, we obtain $\mu \approx 0.1$~ keV, as required to lead to light neutrinos at the sub-eV scale. We want to remark that there is an astonishing accomplishment behind this simple reasoning. Namely, the dependence of $\mu$ on $v$ and $M$ as stated above implies that light neutrino masses can be automatically expressed as $m_\nu\approx \frac{v^2_w}{M}$, which is exactly the mass expression obtained in the canonical see-saw mechanism. In other words, by devising a scheme to provide a naturally small $\mu$ parameter in the ISS, we regain the ordinary relation in the canonical see-saw mechanism.

Finally, there is an amazing feature behind the approach we present here for the ISS  mechanism that we gained for free.  It turns out that $v$, which establishes the mass scale of the RH neutrinos and their mixing with the active ones, and consequently its implications to the nonunitarity effects~\cite{nonunitarityeffect} and lepton flavor violation (LFV) processes~\cite{LFV}, does not play any role in the final expression for the light neutrino masses. Hence, it could lie not only at the TeV scale but even below, which may result in  RH neutrinos popping up at the electroweak scale in LHC. Besides, such a low scale should lead to an enhancement of the non-unitarity effects and LFV processes, turning neutrino experiments even more sensible to the features of RH neutrinos. This is possible, of course, as long as we preserve the hierarchy previously assumed that $\mu << M_D << M_N$. We next show a possible way of realizing such a scheme.

\section{A realistic scenario}
\label{sec:realisticscenario}
We now  present a simple  extension of the SM  which realizes the ISS mechanism proposed here. For this, we increase the SM particle content by adding to it three RH neutrinos: $\nu_{i_R}$, partners of the known LH ones;  three new RH neutral fermions, $N_{i_R}$;  and two neutral scalar singlets $\sigma^0_1$  and $\sigma^{0}_2$. Moreover, we impose that the Lagrangian be invariant  under  a $Z_5\otimes Z_2$ symmetry, with the field's transformation properties given in Table \ref{t1}, where $H$ stands for the usual SM Higgs doublet.  All the SM fermion doublets transform trivially under the discrete symmetry in order to preserve the standard Yukawa Lagrangian for charged fermion mass generation.
\begin{table}
\begin{tabular}{|c|c|c|c|c|c|c|c|c|}
\hline
 & $\nu_{iR}$ & $N_{iR}$ & $\sigma_{1}$ & $\sigma_{2}$ & $H$ & $e_{iR}$ & $u_{iR}$ & $d_{iR}$ \tabularnewline
\hline
\hline
$Z_{5}$ & $1$ & $2$ & $3$ & $4$ & $1$ & $4$ & $4$ & $1$\tabularnewline
\hline
$Z_{2}$ & $0$ & $1$ & $1$ & $0$ & $0$ & $0$ & $0$ & $0$\tabularnewline
\hline
\end{tabular}
\caption{Fields transformation properties under $Z_5\otimes Z_2$. The fermion doublets transform trivially under the discrete symmetries.}
\label{t1}
\end{table}

In this way, the Yukawa interactions allowed by the $Z_5\otimes Z_2$ symmetry are composed by the following terms: 
\begin{eqnarray}
{\cal L}_Y={\cal L}^{{SM}}_Y+ G_{ij}\overline{\nu_{iR}} \tilde{ H}^\dagger L_{jL} +G^\prime_{ij}\sigma_1 \overline{ N_{iR}} \nu^c_{jL} +\frac{\lambda_{ij}}{2} \sigma_2 \overline{N_{iR}} N^c_{jL} + \mbox{H.c}\,,
\label{yukawaSM}
\end{eqnarray}
where $\tilde{H}\equiv \epsilon H^*$, with $\epsilon$ the $SU(2)$ antisymmetric second rank tensor, and ${\cal L}^{{SM}}_Y$ is the SM Yukawa Lagrangian. Although the couplings $G_{ij}$ and $G^\prime_{ij}$ are allowed to be nondiagonal, we assume here that  $\lambda$  is a diagonal matrix, which is the equivalent choice we made for $\mu$ in Sec. (\ref{sec:newmechanism}). We observe that the above Yukawa Lagrangian is automatically invariant under $U(1)_L$, the lepton number symmetry, and this result is a key point to engender the ISS mechanism we have in mind. Shortly, the imposed discrete symmetries do not allow for any explicit lepton number violation at the Yukawa Lagrangian, but we still have the freedom to violate it at the potential, as we are going to see soon.

When the neutral components of $H$,  $\sigma_1$,  and $\sigma_2$ develop VEVs  and are shifted in the usual way,
\begin{eqnarray}
 H^0 , \sigma_1 , \sigma_2 \rightarrow \frac{1}{\sqrt2} (v_{w ,1 ,2}
+R_{ _{w ,1 ,2} } +iI_{_{w ,1 ,2} })\,,
\label{vacua}
\end{eqnarray}
the Yukawa Lagrangian in Eq.~(\ref{yukawaSM}) leads to the same mass matrix for the neutrinos as in Eq.~(\ref{ISSmatrix}) with the same basis $(\nu_L\,,\, \nu^c_L\,,\,N^c_L)$. But now the submatrices composing $M_\nu$ in Eq.~(\ref{ISSmatrix}) are given by
\begin{eqnarray}
& & m_D=\frac{G}{\sqrt{2}}v_w\nonumber \\
& & M_N=\frac{G^{\prime}}{\sqrt{2}}v_1\nonumber\\
& & \mu=\frac{\lambda}{\sqrt{2}} v_2
\label{masstoymodel}
\end{eqnarray}
Thus, for  $\mu << m_D << M_N$, its  diagonalization yields the following effective Majorana mass matrix for the light neutrinos,
\begin{equation}
m_\nu = m_D^T M_N^{-1}\mu (M^T_N)^{-1} m_D=\frac{1}{\sqrt{2}}G^T(G^{\prime})^{-1} \lambda (G^{\prime T})^{-1}G \frac{v^2_w v_2}{v_1^2}\,.
\label{inversenew}
\end{equation}
The order of magnitude of $m_\nu$ in Eq.~(\ref{inversenew}) is determined by $m_\nu \approx \frac{v^2_w v_2}{v^2_1}$.  For $v_w=246$~GeV and $v_1=10^3$~GeV, we need $v_2 \approx 10^{-7}$~GeV to generate neutrino masses at the sub-eV scale.

We now turn to the main feature of the ISS mechanism proposed here. We wish to show that such a small $v_2$ is a consequence of the pattern we choose for the spontaneous breaking of the gauge symmetry. In other words, given the scalar content, $H$, $\sigma_1$,  and $\sigma_2$, with the above assigned vacuum alignment and the additional  $Z_5\otimes Z_2$ symmetry, we are able to obtain a scalar potential with explicit lepton number violation whose minimum constraint equations imply the smallness of $v_2$. The most general potential obeying these requirements can be written as,
\begin{eqnarray}
&V&= \mu^2_H\vert H\vert^2 +\mu^2_1 \vert\sigma_1\vert^2 + \mu^2_2 \vert\sigma_2\vert^2 + \lambda_1 \vert H\vert^4+\lambda_2\vert\sigma_1\vert^4 +\lambda_3 \vert\sigma_2\vert^4 \nonumber \\
&&+\vert H\vert^2(\lambda_4 \vert\sigma_1\vert^2 +\lambda_5\vert\sigma_2\vert^2) + \lambda_6 \vert\sigma_1\vert^2\vert\sigma_2\vert^2
-(\frac{M}{\sqrt2}\sigma_1^2 \sigma_2 + \mbox{H.c}).
\label{potentialSM}
\end{eqnarray}

The last  term of this potential is the one that explicitly breaks the lepton number symmetry.
In substituting the expansions in Eq. (\ref{vacua}) into the potential above, we obtain the following set of tadpole constraint equations:
\begin{eqnarray}
&& \mu^2_H+\lambda_1 v^2_w +\frac{\lambda_4}{2}v_1^2 +\frac{\lambda_5}{2} v_2^2=0,\nonumber \\
&&\mu^2_1+\lambda_2 v^2_1 +\frac{\lambda_4}{2} v_w^2 +\frac{\lambda_6}{2} v_2^2-Mv_2=0,\nonumber \\
&& \mu^2_2 v_2 + \lambda_3 v^3_2 +\frac{ \lambda_5}{2}v^2_w v_2+\frac{\lambda_6}{2}v^2_1 v_2-\frac{M}{2}v^2_1=0.
\label{constraintSM}
\end{eqnarray}

Let us now assume that lepton number is violated at a very high energy scale characterized by $M$ (possibly a grand unified theory scale) in the potential in Eq.~(\ref{potentialSM}). When we consider that  $\sigma_2$ belongs to this high energy scale, which means that  $\mu_2 \approx M$, and for couplings of order of unity, the last term in the set of  constraints  in Eq.~(\ref{constraintSM}) provides,
\begin{equation}
v_2 \approx \frac{1}{2}\frac{v^2_1}{M}.
\label{seesawSM}
\end{equation}

What is amazing here is that with  the relation for $v_2$ given above, the general  expression for the masses of the light neutrinos, Eq.~(\ref{inversenew}) is given by 
\begin{equation}
m_\nu =\frac{1}{2\sqrt{2}}G^T(G^{\prime})^{-1} \lambda (G^{\prime T})^{-1}G  \frac{v^2_w}{M}\,,
\label{typeIseesaw}
\end{equation}
and if we focus only in its order of magnitude~\cite{comment} we have that $m_\nu \approx \frac{v^2_w}{M}$, which is  exactly the same expression that we would obtain for the neutrino masses in the canonical  see-saw mechanism.  Thus, as $v_w$ is the electroweak scale, we need $M=10^{14}$~GeV so as to have light neutrinos at the sub-eV scale. In this way, we now have a deeper understanding that the ISS mechanism actually develops the profile of the canonical see-saw mechanism with neutrino masses being suppressed by a high energy scale.  Moreover, we perceive that  $v_1$, which establishes the mass scale of the RH neutrinos, does not play any role in the final expression of the light neutrino mass. Consequently, we could in principle diminish its value as much as nonunitarity effects and LFV processes allow us to do. This freedom implies that RH neutrinos can be light enough, perhaps even with a mass at the electroweak scale and producible on shell at LHC, with a considerable amount of missing energy as its signature. The phenomenology of a TeV RH neutrino in an  ISS mechanism was explored in Refs.~\cite{probedLHC,nonunitarityeffect,LFV,phenomenology}, and it would be similar here, unless we can really pull the RH neutrino mass down to a few hundreds of GeV, obtaining a further enhancement in the nonunitarity of the PMNS mixing matrix. However, it is opportune to remember that whatever the RH neutrino mass, we have to keep the hierarchy $\mu << m_D << M_N$ in order to make the whole scheme work. This makes it obvious that a careful treatment of $m_D$ should be carried out if we wish to pursue the implications of a hundred-GeV RH neutrino.

Concerning the potential in Eq.~(\ref{potentialSM}), its stability  can be verified straightforwardly by means of mass matrices for the scalar fields. Three CP even scalars arise as combinations of the real parts of $H^0$, $\sigma_1$ and $\sigma_2$, whose quadratic masses are the eigenvalues of the matrix
\begin{equation}
M_R^2=
\begin{pmatrix}
2\lambda_1v_w^2 & \lambda_4v_w v_1 & \frac{\lambda_5}{2}\frac{v_w v_1^2}{M} \\
. & 2\lambda_2v_1^2 & \frac{\lambda_6}{2}\frac{v_1^3}{M} -Mv_1\\
. & . & \frac{\lambda_3}{2} \frac{v_1^4}{M^2}+M^2
\end{pmatrix},
\label{mr}
\end{equation}
in the basis  ($R_w$, $R_1$, $R_2$), where the conditions in Eq.~(\ref{seesawSM}) were taken into account. It can be verified that there are an infinite set of parameters (not fine-tuned) leading to positive eigenvalues. One of these eigenvalues corresponds to an eigenstate identified as the standard Higgs, and the other two being mainly singlets which do not  couple directly to the SM particles.  For the CP odd scalars, it has to be observed that there is an independent global symmetry in Eq.~(\ref{potentialSM}), under which the scalar singlets transform as $\sigma_1\rightarrow e^{-i\alpha}\sigma_1$,  $\sigma_2\rightarrow e^{2i\alpha}\sigma_2$. Once such a symmetry is spontaneously broken an additional  Goldstone boson arises in the spectrum. Thus, the mass matrix for  the CP odd fields furnishes two zero eigenvalues, with one of them being the eigenstate which forms the longitudinal component of the standard $Z^0$ gauge boson, and the other one being an eigenstate which is a combination of imaginary components of $\sigma_1$ and  $\sigma_2$. This last Goldstone boson does not represent a threat to the model because it is a singlet by the SM symmetry, and  its main  impact would be on the decay of heavy new neutrinos into lighter ones. Also, there is a heavy CP odd eigenstate with mass $m_A=\sqrt{v_1^2+M^2}$. All of this shows us that the potential has a minimum consistent with spontaneous symmetry breaking for the VEVs we assumed here.

\section{Conclusions}
\label{sec:conclusions}
We have presented a development for the ISS mechanism considering that lepton number is violated explicitly through the scalar potential of a simple extension of the SM. The scale responsible for inducing a tiny mass for the light neutrinos, the parameter $\mu$ in the ISS mechanism, is intimately connected to lepton number violation and necessarily assumed to be of the order of keV. While it can be viewed as a naturally occurring scale in the sense that it mildly breaks the lepton number symmetry, its smallness was lacking a theoretical explanation. To build a minimal extension of the SM to enlighten this question was the goal of this work. We did that by introducing the usual six extra neutral fermions to engender the ISS mechanism, as well as two extra singlet scalar fields, along with a $Z_5\otimes Z_2$ symmetry. We have shown that the canonical see-saw formula is obtained in our scheme in a way that is independent of the RH neutrino mass scale, which may be at the electroweak scale and accessible at LHC and also lead to enhanced effects concerning the nonunitarity of the PMNS mixing matrix.

\acknowledgments
This work was supported by Conselho Nacional de Pesquisa e
Desenvolvimento Cient\'{i}fico- CNPq.  A.G.D. also thanks FAPESP for supporting this work.



\begin{thebibliography}{99}
\bibitem{comment}
 It is not necessary to go into details about the exact values of the couplings, although they can naturally be taken to be $\cal{O}$(1) in our scheme.
\end{thebibliography}
\end{document}